\documentclass[letterpaper,twocolumn,10pt]{article}
\usepackage{usenix-2020-09}

\usepackage[normalem]{ulem}
\usepackage{fdsymbol}
\usepackage{algorithmic}
\usepackage{graphicx}
\usepackage{textcomp}
\usepackage{xcolor}
\usepackage{svg}
\usepackage{amsmath}
\usepackage{caption}
\usepackage{comment}
\usepackage{subcaption}
\usepackage{soul}
\usepackage{booktabs}

\begin{document}
\author{\normalfont
  \begin{tabular}{c @{\hskip 1.75in} c}
    \begin{tabular}{c}
    {Shivank Garg}\thanks{Work done during his internship at Intel Labs, Bengaluru}\\
    IIT Kanpur\\
    Kanpur, India
    \end{tabular}
 &
    \begin{tabular}{c}
    {Aravinda Prasad}\\
    Processor Architecture Research Lab\\Intel Labs, India
    \end{tabular}
    \\[0.85cm]
    \begin{tabular}{c}
    {Debadatta Mishra}\\
    IIT Kanpur\\
    Kanpur, India
    \end{tabular}
 &
    \begin{tabular}{c}
    {Sreenivas Subramoney}\\
    Processor Architecture Research Lab\\
    Intel Labs, India
    \end{tabular}
    \end{tabular}
    }
\date{}

\title{Motivating Next-Generation OS Physical Memory Management \\ for Terabyte-Scale NVMMs}

\maketitle

\begin{abstract}

Software managed byte-addressable hybrid memory systems consisting of DRAMs and NVMMs offer a lot of flexibility to design efficient large scale data processing applications.
Operating systems (OS) play an important role in enabling the applications to realize the integrated benefits of DRAMs' low access latency and NVMMs' large capacity along with its persistent characteristics.
In this paper, we comprehensively analyze the performance of conventional OS physical memory management subsystems that were designed only based on the DRAM memory characteristics in the context of modern hybrid byte-addressable memory systems. 

To study the impact of high access latency and large capacity of NVMMs on physical memory management, we perform an extensive evaluation on Linux with Intel's Optane NVMM.
We observe that the core memory management functionalities such as page allocation are negatively impacted by high NVMM media latency, while functionalities such as conventional fragmentation management are rendered inadequate. We also demonstrate that certain traditional memory management functionalities are affected by neither aspects of modern NVMMs. 
We conclusively motivate the need to overhaul fundamental aspects of traditional OS physical memory management in order to fully exploit terabyte-scale NVMMs. 

\end{abstract} 

\section{Introduction}
\label{sec:introduction}
The amount of digital data generated, processed and stored is exponentially
increasing for the modern data intensive workloads such as 
big data analytics, graph processing, DNNs.
Furthermore, DRAM memory capacity is not scaling at a proportional
rate~\cite{dram_scale_1,dram_scale_2,dram_scale_3} resulting in exploration 
of several new hardware memory technologies with diverse capabilities~\cite{korgaonkar2018density, kultursay2013evaluating, 8649762}.
Intel’s Optane DC PMM nonvolatile main memory (NVMM)~\cite{optane_dc_briefs}
is one such emerging memory technology that 
has the potential to cater to the demanding memory capacity 
of data intensive applications.
To satisfy the large scale memory capacity requirements of data intensive workloads, 
enterprise datacenters as well as cloud service providers are building hardware platforms 
that increasingly combine traditional DRAM memory technology along
with emerging NVMM memory technologies such as Optane~\cite{8649762, baidu}.
The diverse memory tiers are typically exposed as a unified byte-addressable
main memory to the software for better efficiency, utilization and flexibility.

\begin{figure}[tp]
\centering
\includegraphics[width=0.95\linewidth]{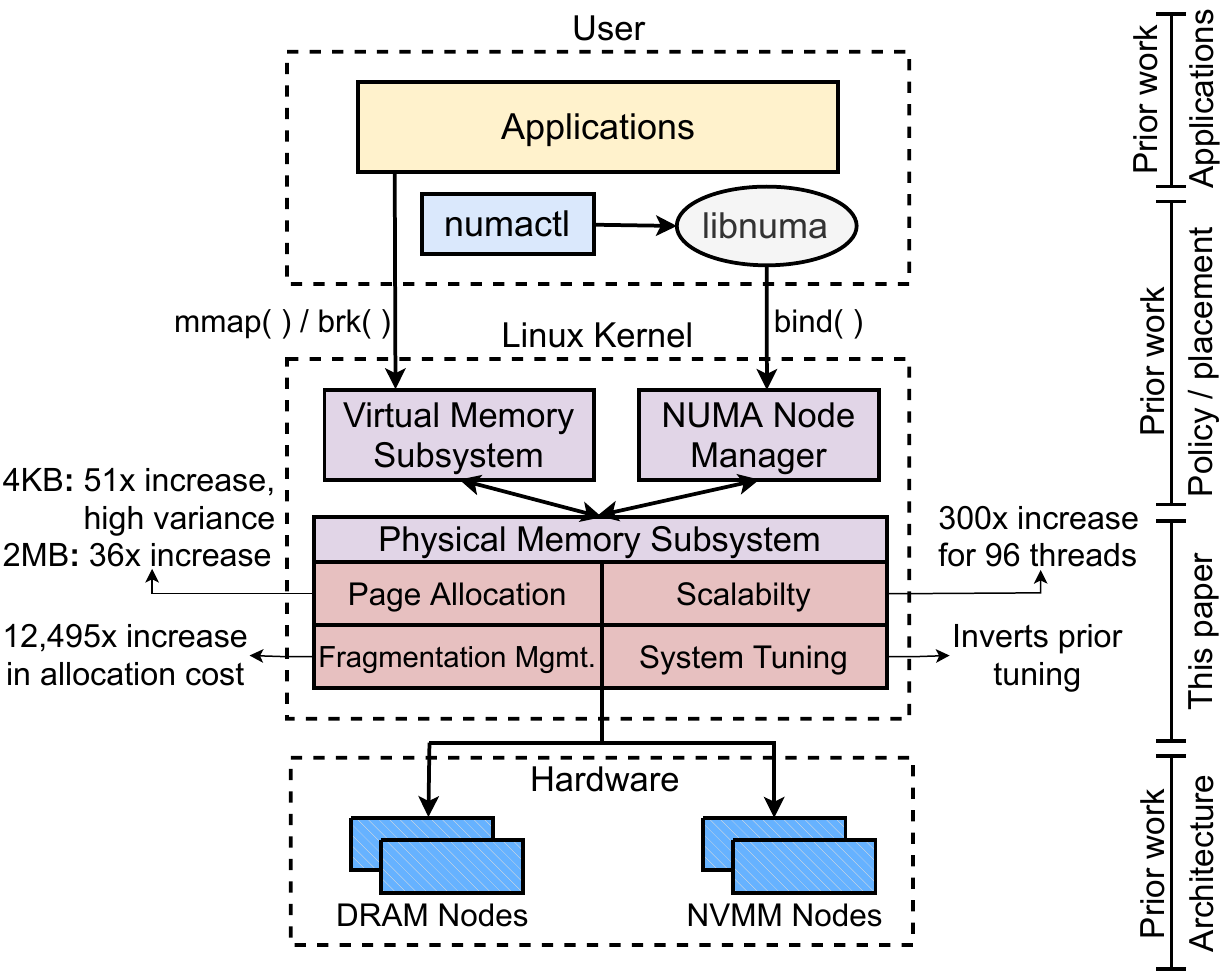}
\caption{Scope and summary of our characterization work. The diagram also depicts the  hybrid memory configuration and usage of DRAM and NVMM as NUMA nodes.}
\label{fig:numa_flat_mode}
\end{figure}

Techniques at different layers of the software stack try to 
take advantage of the heterogeneous memory by using its 
unique characteristics in different ways.
Efficient in-memory persistent data structures~\cite{NVheaps,mnemosyne,PersistentBtreeNVM}, 
placement of data across the memory tiers~\cite{nimble,bridge_dram_nvmm,memos,heteroos,kumar2021radiant},
new or customized abstractions for NVMM access~\cite{Bittmantwizzler,nova, splitfs,hotos13NVMabs, BittmanNVMpointers,BittmanTaleof2abs,dulloorNVMsystemsoftware}
are some interesting research directions leveraging the exposure of 
NVMM to the software layers.
While all of the above research directions are important, the core OS memory 
management routines that manage the memory resources 
for any computer system play a significant role.
The {\em effectiveness of the OS physical memory management techniques} 
for tiered or hybrid memory
systems is the scope of this paper (Figure~\ref{fig:numa_flat_mode}).

The techniques used in many production grade OSes
are designed and optimized based on DRAM as the underlying memory 
hardware. 
As it has been a case in the past, any hardware change (e.g., NUMA systems~\cite{autonuma, numactl}, Huge page support~\cite{hugethpsupport}) required 
design changes of different magnitudes at the system software layer
for improved efficiency.
In case of hybrid memory systems, the transition is non-trivial as DRAMs and NVMMs differ 
both in terms of their access latencies and memory capacity.

In this paper, we analyze the efficacy of widely accepted
design artifacts of OS physical memory management in the context of  
hybrid memory systems by answering the following fundamental questions.
What are the performance and scalability requirements on core OS physical memory management functionalities warranted by modern NVMMs?
Considering the large capacity and comparatively high access latency of NVMMs, do the physical memory management optimizations perform as well as systems with only DRAM?

We answer the above questions by characterizing 
the Linux operating system on a server configured with 
Intel's Optane DC PMM nonvolatile memory. 
We consider the core functionalities such as physical memory bookkeeping and management along with physical memory allocation/deallocation in this study.
We design a microbenchmark suite containing specialized 
in-kernel workloads to invoke memory management functions in a 
controlled manner with different parameters for our performance analysis. 
To precisely identify performance bottlenecks
we instrument the kernel to capture additional 
performance statistics at a fine granularity. 

Based on extensive analysis and insights into the system performance and scalability  (Figure~\ref{fig:numa_flat_mode} highlights the summary of our characterization work), 
we argue that many traditional memory management optimizations that were designed for DRAM-only systems are often insufficient and sometimes degrade system performance when NVMMs are introduced. 
We then propose multiple key recommendations towards an optimized overhaul of OS physical memory management for hybrid memory systems: 

\begin{itemize}
    \item Avoid page zeroing from the OS page allocator to exploit NVMM's capacity advantages.
    \item Aggressively employ fragmentation avoidance techniques to eliminate exorbitantly high allocation latency on fragmented high capacity NVMMs.
    \item By showing that page allocation on NVMMs is sensitive to interference from other memory management activities, we recommend carefully coordinated techniques across memory management functionalities.
\end{itemize}

Furthermore, in \S\ref{sec:discussions}, we explain how memory-type agnostic system-wide memory management schemes are sub-optimal in tiered memory systems, and argue for differential memory management for distinct memory tiers. 
We also motivate in detail the need to develop novel memory management techniques for terabyte scale NVMMs.

\section{Background and Motivation}
\subsection{Intel's Optane DC PMM}
Intel's Optane DC PMM is an emerging high capacity nonvolatile main memory that is
affordable and flexible (supports both volatile and non-volatile modes).
Optane\footnote{Optane and NVMM are used in an interchangeable manner in this paper.} is DDR4 socket compatible and can be plugged into conventional DRAM DIMM slots~\cite{optane_dc_briefs, optane_dc_quickstart}.
Optane DIMMs offers high capacity supporting up to 24\,\text{TB} of main memory~\cite{optane_dc_briefs} and has lower cost per byte in comparison to DRAM, but incurs higher access latency (typically 2$\times$--3$\times$ higher than DRAM ~\cite{fastpaper}).
Optane can be configured either as a byte-addressable volatile main memory (Memory Mode or Flat Mode) or a block-addressable non-volatile memory (App Direct Mode)
using tools and utilities provided in Persistent Memory Development Kit (PMDK)~\cite{PMDK_home}.

\noindent
{\bf Memory Mode:}
In this mode Optane is configured as a byte-addressable main memory while DRAM acts as a large L4 cache. Any data accessed on Optane is moved to DRAM by the hardware at a cache line granularity. This mode enables easy adoption of large capacity Optane 
as it does not require any application or OS changes.
Due to lack of any flexibility in the software in managing or placing data across memory tiers we do not consider this mode for our characterization.

\noindent{\bf Flat mode:}
In this mode both DRAM and Optane DIMMs are visible to the software and can be accessed as a unified and heterogeneous main memory as shown in Figure~\ref{fig:numa_flat_mode}. 
Due to data management and placement flexibility provided in this mode,
software including the OS plays a significant role. Hence, we consider this mode for our evaluation. 

\subsection{Tiered Memory Systems}
The tiered memory systems configured in Flat mode enable software to exploit the best of different heterogeneous memories~\cite{nimble}. For instance, NUMA systems~\cite{numa_docs} comprising of the low-latency DRAM and the high-capacity NVMM can efficiently cater to the needs of modern data centers~\cite{dulloor_hetero_system,exploitnumaassymetry}.
As shown in Figure~\ref{fig:numa_flat_mode}, applications can use NUMA related system calls~\cite{numa7} or \texttt{libnuma}~\cite{numa3} library for configuration and tuning. 
Moreover, Linux provides {\tt numactl}~\cite{numactl}, a command-line utility, to simplify 
the node binding process without any application changes.
The physical memory management subsystem allocates pages based on the 
binding requests from the applications. 

\subsection{Linux Physical Memory Management}
Efficient memory management plays a crucial role in system performance.
An ideal memory allocation mechanism should be able to allocate group of contiguous pages with minimum latency while keeping fragmentation in check~\cite{gorman2006and}.

\begin{figure}[tp]
\centering
\includegraphics[scale=0.45]{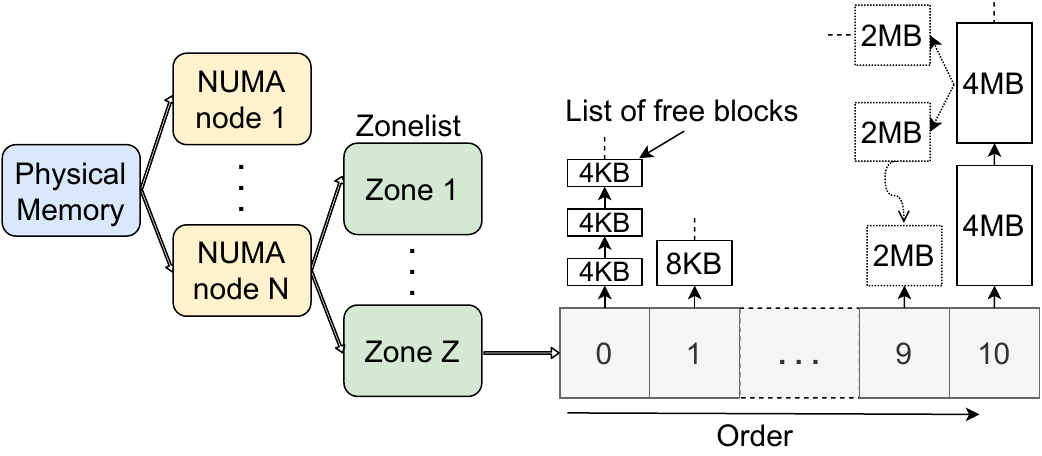}
\caption{Physical memory management using buddy allocation
     technique in the Linux kernel.}
\label{fig:linux_buddy_alloc}
\end{figure}

The Linux kernel uses binary buddy allocation technique for page allocation and freeing~\cite{buddy_alloc1},
and employs heuristics to minimize memory fragmentation~\cite{gorman2004understanding}.
Physical memory is abstracted as NUMA nodes where each NUMA node is divided into separate memory ranges called {\tt zones}.
Each zone maintains a linked list of free blocks of different orders where each block is of size \mbox{2\textsuperscript{order} $\times$ 4\,KB} as shown in  Figure~\ref{fig:linux_buddy_alloc}. When an application requests for physical memory, the buddy allocator traverses through the list of zones in each permitted NUMA node (giving priority to the node binding requests from the application and local node in that order) to find a free physical memory region. 
If a free block of requested size is not available then the buddy allocator splits a higher order block into two `buddies' where one of the buddies is added to the lower order free list while the other block is split further until the desired block size is reached.
During freeing, the buddy allocator repeatedly merges the pair of free buddies (if available) to form a higher order page block.

\subsection{Huge Page Support}
Huge pages reduce address translation overheads for large memory systems~\cite{hugepages1_lwn,ingens,redhatthptuning} mainly due to--- (i) increased TLB coverage (less number of TLB misses), and (ii) reduced cost of page table traversal in case of a TLB miss. Many architectures support multiple huge pages, for example, x86 supports 2\,\text{MB} and 1\,\text{GB} pages besides standard 4\,\text{KB} pages. 

Transparent Huge Pages (THP)~\cite{hugethpsupport} provides developer-friendly transparent support to applications for using 2\,\text{MB} huge pages.
THP can be configured and tuned using \texttt{sysfs} parameters~\cite{hugethpsupport}.
For example, a user can enable or disable THP support completely, or can enable it selectively for memory regions using \texttt{madvise}~\cite{madvise} system call.
THP defragmentation configuration allows the user to either allocate huge pages immediately or defer it. In the first case, if continuous memory region is not available, the application will stall and compact memory which may impact latency.
In the latter case, application is allocated regular pages which are promoted to THP later by memory management daemons.

\subsection{Motivation}

Physical memory management is an important part of memory management, especially for terabyte-scale Optane memory. A higher page allocation latency can create amplified adverse impact hurting startup and runtime performance of applications. Furthermore, many large memory footprint applications spend significant part of the execution time in the kernel and we observe that the time spent in kernel increases for Optane memory. 
For example, BFS, a graph analytics workload spends around 15.1\% of the total execution time in kernel with DRAM memory which increase to 26.6\% with Optane memory. 
Earlier works have looked into characterizing the performance of Optane at architectural~\cite{fastpaper} and application level~\cite{optane_graph,10.1145/3451342,hpc_optane_measurement}.
However, they have not focused on the implications of physical memory management at terabyte scale.
Hence, it is imperative to characterize the performance of critical memory management functionalities such as page allocation on Optane memory.

\section{Experimental Setup}
\begin{figure}[tp]
\centering
\includegraphics[scale=0.9]{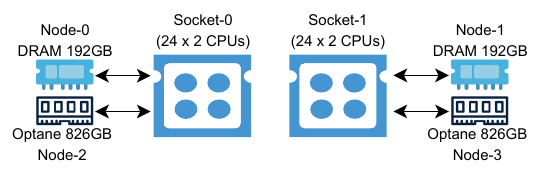}
\caption{A 2-Socket NUMA system used for our evaluation}
\label{fig:NUMA_optane_system}
\end{figure}

We use Intel-Xeon Gold 6252N with 96 CPUs (2 sockets, 24 cores, 2-way HT), 192\,\text{GB} DRAM and 826\,\text{GB} Optane connected to each socket (Figure~\ref{fig:NUMA_optane_system}). We use PMDK~\cite{PMDK_home} to configure our system in Flat mode. PMDK registers Optane-backed NUMA nodes as no-CPU NUMA nodes while DRAM NUMA nodes have 48 CPUs each. 
We use Fedora 30 with 5.6.13 Linux kernel. 
We use 4-level paging as it can map up to 256\,TB physical memory, disable address space layout randomization (ASLR) feature, and set DVFS to performance. 

\subsection{Benchmark Infrastructure}

We mainly use microbenchmarks to characterize physical memory management because:
(i)~workloads exploiting NVMMs are increasingly getting diverse from traditional OLTP and HPC workloads to modern big data, AI/ML and cloud-native workloads~\cite{hpc_optane_measurement, optane_graph, autotm, baidu, bridge_dram_nvmm, heteroos, panthera, 10.1145/3451342}. Thus selecting a representative set of workloads is challenging, (ii)~only certain phases of the workloads are sensitive to physical memory management performance. For example, page faulting on virtual memory region without a backing physical page when initializing a large working data set, performing hot/cold memory rebalancing across tiers and during large page promotions/demotions. But in phases of the workloads where physical memory management is not actively involved, the performance of the application depends on NVMM access frequency (which we do not intent to characterize). It is thus challenging to capture, analyze and associate performance bottlenecks to memory management functionalities during the execution of real-world benchmarks, (iii)~microbenchmarks help in understanding the worst-case performance scenarios
which are important for tail latency analysis and for determining service level objectives (SLOs). 

We design a microbenchmark infrastructure that consists of kernel modules and userspace programs/utilities.
We experiment with different parameters including
memory size, target NUMA node(s), page order, number of concurrent threads and CPU bindings.
System configuration, benchmark methodology for characterizing individual functionality are explained in detail in the characterizing section (\S\ref{characterization}).

To measure and analyze performance statistics at a fine granularity we instrument the kernel by adding custom trace points. We also capture relevant memory management statistics available in \texttt{/proc}.

\section{Characterization}
\label{characterization}

In this section, we present the efficacy of the Linux kernel's memory management with Intel's Optane DC PMM.

\subsection{Page allocation}
\label{char:page-allocation}

\begin{figure*}
            \begin{subfigure}[h]{0.48\linewidth}
            \includegraphics[width=\linewidth]{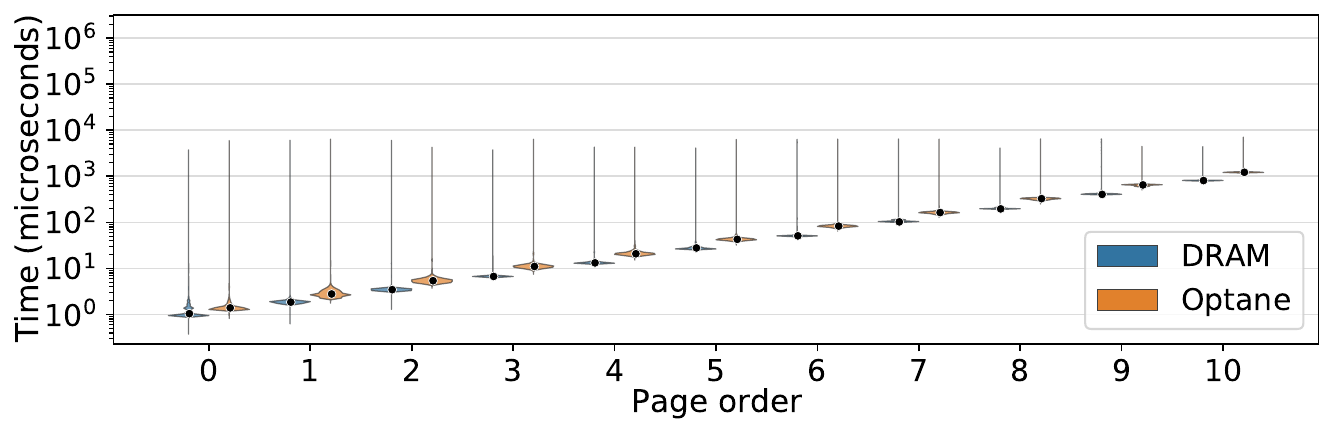}
            \caption{Isolated page allocation}
            \label{fig:vio_alloc_cost_1thread_DRAMOPTANE}
            \end{subfigure}
            \begin{subfigure}[h]{0.48\linewidth}
            \includegraphics[width=\linewidth]{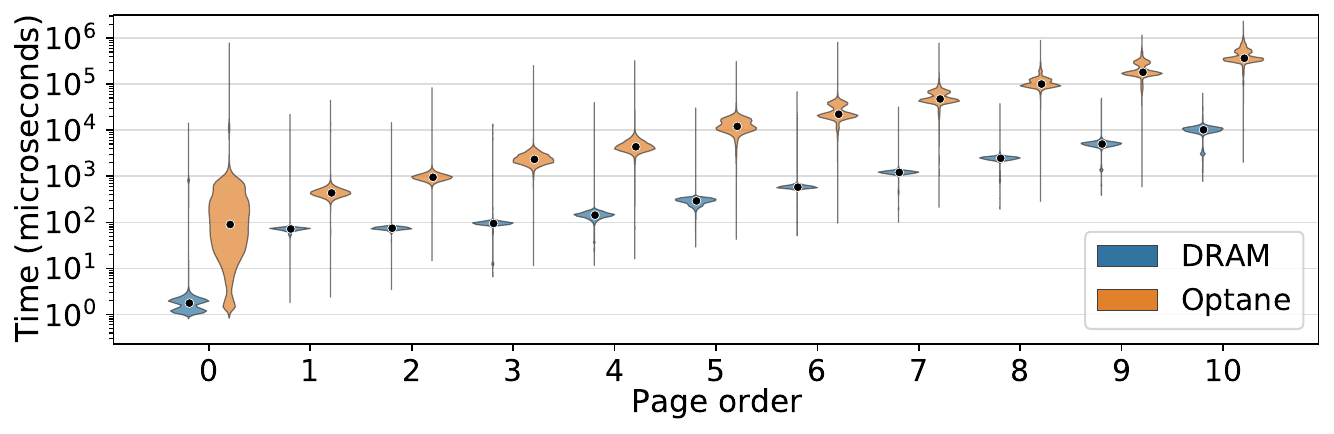}
            \caption{Concurrent page allocation}
            \label{fig:vio_alloc_cost_96thread_DRAMOPTANE}
            \end{subfigure}
    \caption{Page allocation cost for different page orders on DRAM and Optane. The violins show the distribution of the allocation cost, the $\bullet$ represents the median and the whiskers show the minimum and maximum values. The y-axis is set to log scale.}
    \label{fig:vio_alloc_cost_DRAMOPTANE}
\end{figure*}
Allocating a page is the frequently invoked, performance sensitive and critical functionality of the operating system's physical memory management subsystem.
A page is allocated when a process faults on a valid virtual address region without a backing physical frame. Hence, the time spent in allocating a page frame directly impacts the page fault latency which in turn impacts application's performance and tail latency. 
In addition, many Linux kernel subsystems such as the SLAB allocator directly invokes the buddy allocator to allocate pages. 
    
We characterize the page allocation cost for different page orders on DRAM and Optane. For each page order, we directly invoke the interface exported by the Linux kernel's buddy allocator in an open loop from our benchmark kernel module to allocate a total of 150\,GB of memory on either DRAM or Optane NUMA nodes. 

We measure the page allocation overheads in two scenarios (i) isolated scenario, where we invoke one instance of page allocation at a time on an idle system and (ii) stressed scenario where we trigger concurrent page allocations on all the CPUs in the system. Concurrent page allocations are common in applications that support multi-threaded or multi-process initialization of a large working set~\cite{bfs, memcached, redis, SilentShredder}.

    \subsubsection{Isolated allocations}
    \label{char:single-threaded-alloc}

    We observe that the median page allocation cost is higher in Optane compared to DRAM for all page orders (Figure~\ref{fig:vio_alloc_cost_1thread_DRAMOPTANE}).
    The median cost of allocating an order-0 page (4\,\text{KB}) 
    on Optane compared to DRAM
    increases by 33.55\%, while the median cost increases by around 60\% for higher order pages. However, on both DRAM and Optane, the maximum allocation cost is several orders higher than the median cost for most of the page orders.
    
\subsubsection{Concurrent allocations}
\label{char:multi-threaded-alloc}
We observe that the median allocation cost increases by orders of magnitude for Optane compared to DRAM for all page orders (Figure~\ref{fig:vio_alloc_cost_96thread_DRAMOPTANE}). For Optane, the allocation costs have high variance as can be seen from the violin plots. The tail allocation cost for Optane is also significantly higher than DRAM.

The performance of order-0 or 4\,\text{KB} page on Optane is of particular interest as it results in pathological performance (analyzed in detail in the next subsection). The allocation cost is distributed over several orders of magnitude with many allocations costing around 1\,\text{ms} (stretched violin plot in Figure~\ref{fig:vio_alloc_cost_96thread_DRAMOPTANE}) resulting in 51$\times$ increase in median allocation cost compared to DRAM. 

\subsubsection{Discussion}

   Typical Optane media latency is 2$\times$--3$\times$ higher than DRAM~\cite{fastpaper}. Interestingly, the higher Optane access latency is not manifesting itself to page allocation cost in isolated allocation scenario as the allocation cost increased by 33--60\% compared to DRAM. This is due to non-media overheads such as kernel metadata management dominating the allocation cost.
   
   However, the page allocation cost increases significantly (e.g., by 51$\times$ for order-0 page) compared to DRAM in concurrent allocation scenario which is far more than 2$\times$--3$\times$ higher Optane media latency. 
   Prior work on hardware characterization of Optane suggests limiting concurrent access to Optane DIMMs as they cause contention in NVDIMMs internal buffers~\cite{fastpaper}. 
   Even though OSes can control or limit the concurrent page allocations that are triggered by concurrent page faults in applications, doing so will increase page fault latency which negatively impacts application's performance.
   Hence, we believe that novel page allocation optimizations targeted for concurrent page allocations on Optane should be explored.
   
\subsection{Detailed analysis of page allocation overheads}    
\label{char:page_alloc_overheads}

We categorize and measure the page allocation cost into three primary heads: (i) finding a free page, which includes finding the suitable zone
to allocate the page, splitting the higher order pages or merging the lower order pages when required, (ii) zeroing time, the time required to write zeros into the newly allocated page. Clearing or zeroing a page is a {\em security requirement} to avoid information leaks as a freed page may have been used by a different process in the past (iii) other overheads, which include bookkeeping activities and sanity checks.

\begin{figure}
\centering
    \begin{subfigure}[h]{0.45\linewidth}
        \includegraphics[width=\linewidth]{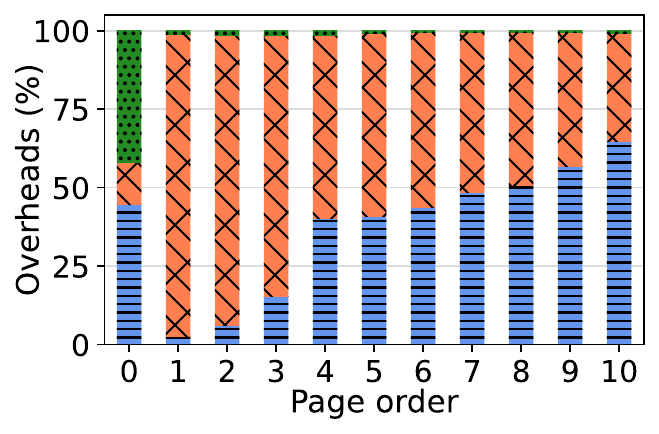}
        \caption{DRAM}
        \label{fig:bar_alloc_cost_96thread_DRAM}
    \end{subfigure}
    \begin{subfigure}[h]{0.45\linewidth}
        \includegraphics[width=\linewidth]{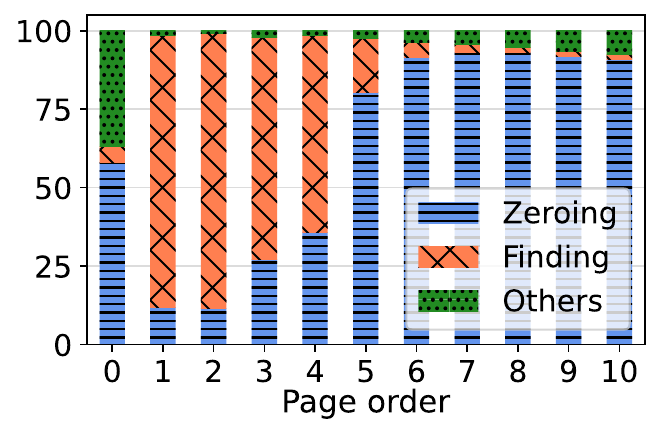}
        \caption{Optane}
        \label{fig:bar_alloc_cost_96thread_OPTANE}
    \end{subfigure}
    \caption{Overheads of page allocation in finding free memory and zeroing the page for different page orders on DRAM and Optane for concurrent allocation scenario with 96 concurrent threads}
    \label{fig:bar_alloc_cost}
\end{figure}

Figure~\ref{fig:bar_alloc_cost} shows the average page allocation overheads for concurrent allocation scenario. It can be observed that the cost of finding a free page is small for order-0 allocations on DRAM and Optane. This is because the kernel uses a per-cpu list of free pages to serve an allocation request where these lists are filled and drained in batches. As a result, allocation requests served from per-cpu lists do not need to acquire zone locks which in turn significantly reduces lock contention thus reducing the cost of finding a free page. However, the cost of finding a free page is significant for all other page orders in DRAM due to contention on zone locks. In case of Optane, although the cost of finding a free page is high for page order~one and above, the zeroing cost starts dominating and accounts for more than 90\% of the overheads for page order~six and above.

\subsubsection{Analyzing performance pathology on Optane for order-0 pages}
\label{char:analyze-performance-overhead-alloc}

    \begin{figure}
    \centering
    \begin{subfigure}[h]{0.45\linewidth}
        \includegraphics[width=\linewidth]{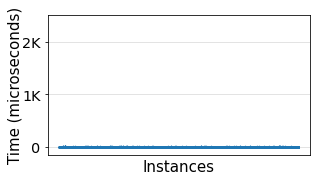}
        \caption{DRAM: Zeroing time}
        \label{fig:scat_zero_0order_96thread_DRAM}
        \end{subfigure}
        \begin{subfigure}[h]{0.45\linewidth}
        \includegraphics[width=\linewidth]{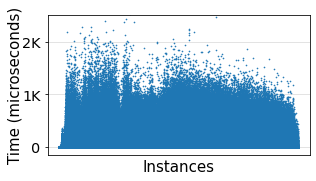}
        \caption{Optane: Zeroing time}
        \label{fig:scat_zero_0order_96thread_OPTANE}
        \end{subfigure}
        \caption{Scatter plot depicting time taken for zeroing a newly allocated page on DRAM and Optane memory in concurrent allocation scenario for page order 0.}
        \label{fig:scat_0order_96thread}
    \end{figure} 

We analyze the pathological performance of order-0 or 4\,\text{KB} page for Optane in concurrent allocation scenario (stretched violin plot for order 0 in Figure~\ref{fig:vio_alloc_cost_96thread_DRAMOPTANE}) with the help of scatter plots shown in Figure~\ref{fig:scat_0order_96thread}. Each dot in the plot represents the overheads associated with an invocation instance of allocating an order-0 page.  

We observe that zeroing time on Optane (Figure~\ref{fig:scat_zero_0order_96thread_OPTANE}) is exorbitantly high for many page allocation instances.
Even though both DRAM and Optane incurs zeroing overheads, the absolute cost of zeroing for DRAM is significantly less than Optane. 
In addition, the zeroing cost on DRAM is mostly contained with a small variance while on Optane it varies significantly.
These observations suggest that zeroing cost significantly increases page allocation time and is the main contributor for the stretched violin plot for order 0 in Figure~\ref{fig:vio_alloc_cost_96thread_DRAMOPTANE}.

\subsubsection{Discussion}

We observe a clear shift in page allocation bottlenecks from DRAM to Optane. While the main bottleneck in DRAM for higher order pages is lock contention during finding a free page, the bottleneck shifts to zeroing a page on Optane. This is due to differences in media access characteristics on DRAM and Optane.

Even though page zeroing is a known problem, many production operating systems have not seriously explored optimizations to eliminate zeroing as it does not significantly contribute to page allocation cost on DRAM. However, our quantification of page zeroing overheads on Optane suggests that page zeroing
overheads cannot be ignored as it significantly contributes to page allocation overheads (more than 90\% for higher order allocations) and also results in high variance in order-0 page allocation time. 

Furthermore, order-0 pages are the most frequently requested pages and thus are optimized using per-cpu lists as mentioned above to keep the allocation cost as low as possible. Clearly, such optimizations on Optane, even though useful, are not sufficient; additional techniques on top of such optimizations are needed to address the overheads on Optane.

\subsection{Page freeing}
\label{char:page-freeing}
\begin{figure}
    \begin{subfigure}[h]{\linewidth}
        \includegraphics[width=\linewidth]{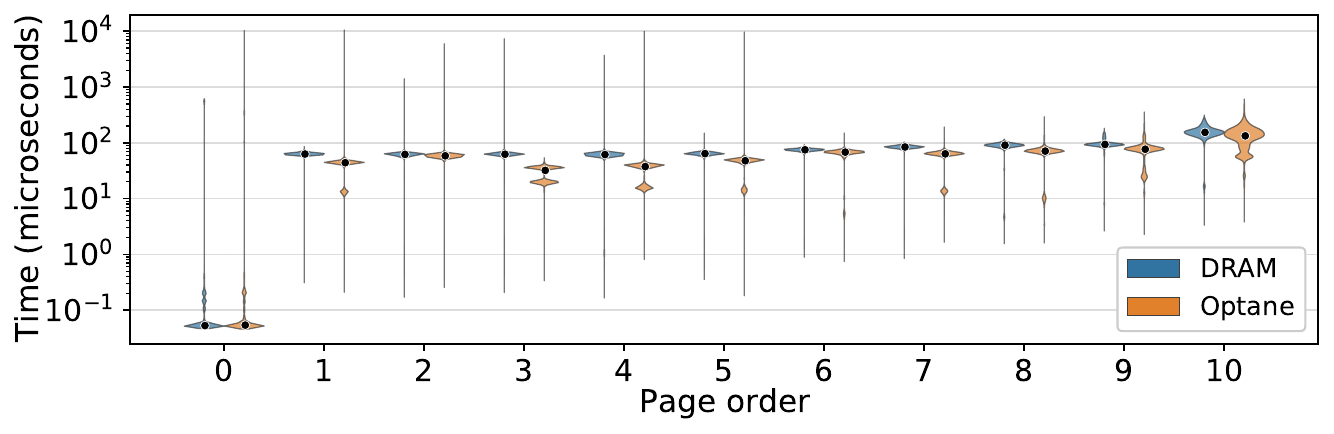}
    \end{subfigure}
    \caption{Page freeing cost for freeing different page orders on DRAM and Optane. The violins show the distribution of the freeing cost, the $\bullet$ represents median and the whiskers show the minimum and maximum value.}
    \label{fig:vio_free_cost_96thread}
\end{figure}

Like page allocation, freeing a page is also a frequently invoked function. When a page is freed, the buddy allocator in the Linux kernel adds the page to the list of free pages maintained per zone (or per-cpu for order-0 pages) and merges the adjacent free pages or buddy pages to form a free list of higher order pages.
To analyze the impact of Optane on page freeing, we allocated and freed 150\,\text{GB} memory for each page order on DRAM and Optane NUMA nodes and measured page freeing time. We present results only for concurrent freeing scenario as the isolated freeing scenario exhibits similar trends. 

The violin plots in Figure~\ref{fig:vio_free_cost_96thread} are almost similar for both DRAM and Optane because freeing a page
does not touch the physical page (unlike page allocation which performs zeroing). Freeing an order-0 page is efficient compared to other page orders due to per-cpu free page list.

\subsubsection{Discussion}

Freeing a page is not impacted by high latency and large capacity of Optane memory. Surprisingly, the cost of freeing a page on Optane is better than DRAM for page order one and higher. Our deeper analysis reveals that the minor improvement in page freeing cost is due to the buddy allocator skipping the coalescing of buddy pages on Optane. The buddy allocator decides to coalesce pages based on the availability of the number of higher order free pages. On Optane, due to the larger memory capacity compared to DRAM, there are several higher order free pages available,  thus the buddy allocator rightfully skips coalescing pages. Such a strategy in the buddy allocator renders adequate for large capacity NVMMs.

\subsection{Memory locality}
\label{char:allocation-locality}
\begin{figure}
    \begin{subfigure}[h]{0.2318\linewidth}
        \captionsetup{justification=centering}
        \includegraphics[width=\linewidth]{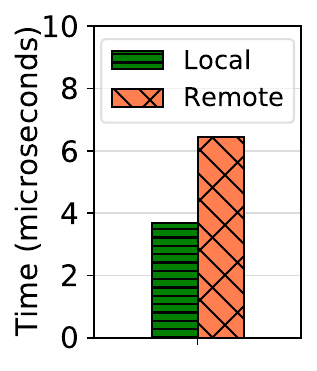}
        \caption{DRAM:\\ \,\,\,\,\,\,\,\,\,Order 0}
        \label{fig:bar_dram_alloc_remote_or0}
    \end{subfigure}
    \begin{subfigure}[h]{0.2508\linewidth}
        \captionsetup{justification=centering}
        \includegraphics[width=\linewidth]{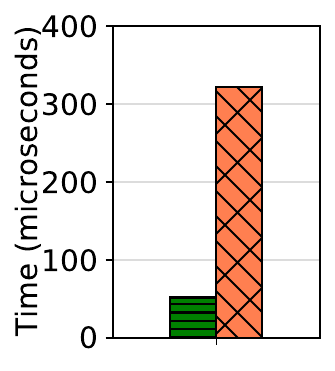}
        \caption{Optane:\\ \,\,\,\,\,\,\,\,\,Order 0}
        \label{fig:bar_optane_alloc_remote_or0}
    \end{subfigure}
    \begin{subfigure}[h]{0.2327\linewidth}
        \captionsetup{justification=centering}
        \includegraphics[width=\linewidth]{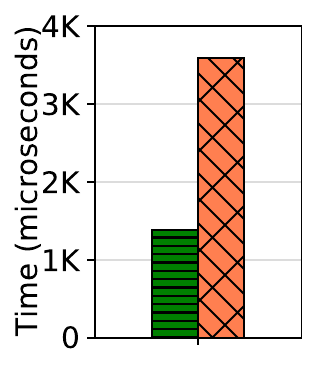}
        \caption{DRAM:\\ \,\,\,\,\,\,\,\,\,Order 9}
        \label{fig:bar_dram_alloc_remote_or9}
    \end{subfigure}
    \begin{subfigure}[h]{0.2607\linewidth}
        \captionsetup{justification=centering}
        \includegraphics[width=\linewidth]{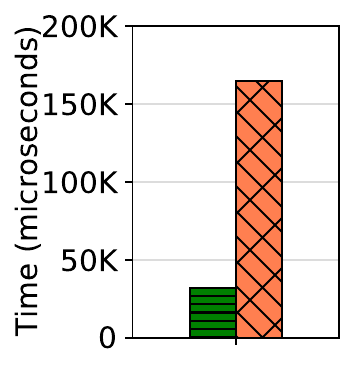}
        \caption{Optane:\\ \,\,\,\,\,\,\,\,\,Order 9}
        \label{fig:bar_optane_alloc_remote_or9}
    \end{subfigure}
    \caption{Average time to allocate order-0 and order-9 pages on local and remote NUMA nodes for DRAM and Optane. Note the difference in y-axis scale}
    \label{fig:bar_alloc_remote}
\end{figure}

Page allocators prefer to allocate pages from the local NUMA node.
However, in many scenarios local allocations are either not possible or not allowed: (i) when interleaved allocation policy is applied, pages are allocated in a round robin manner on both local and remote NUMA nodes~\cite{numa_challenges, traffic_mgmt_numa} and (ii) when free memory is not available in local NUMA node, allocator is forced to allocate from remote NUMA nodes. In this section, we analyze the impact of allocation locality (local or remote NUMA node) on page allocation cost for DRAM and Optane.
We allocate 150\,\text{GB} memory with 48 concurrent threads and bind them to CPUs on either local or remote NUMA nodes for order-0 and order-9 pages.

Figure~\ref{fig:bar_alloc_remote} shows the impact of allocation locality on DRAM and Optane. 
We observe that the page allocation cost significantly increases for remote Optane allocations for both order-0 and order-9 pages. In comparison, DRAM is less affected by NUMA locality of page allocations.

\subsubsection{Discussion}

For existing page allocators designed and optimized based on DRAM as the underlying memory hardware, the cost of allocating a page from remote NUMA nodes, whenever required, is acceptable. 
However, due to costly cross socket allocations with Optane memory, page allocators should efficiently offload remote allocations by having, for example, per-socket page allocation daemons.

Furthermore, using interleaved policy to balance memory allocations across NUMA nodes is not optimal on a tiered memory system as it incurs costly cross socket allocations. Hence, it is worth exploring techniques that effectively balance allocation cost across memory tiers.

\subsection{Multi-threaded scalability}
\label{char:multi-threaded-scalability}

    \begin{figure}
    \centering
        \begin{subfigure}[h]{0.45\linewidth}
        \includegraphics[width=\linewidth]{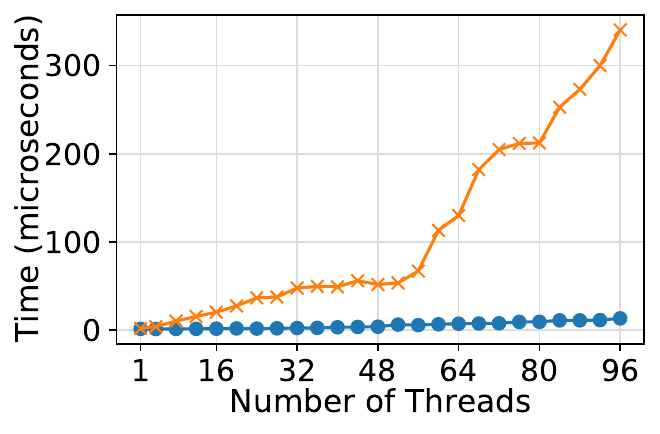}
        \caption{Page order 0}
        \label{fig:line_alloc_varth_or0}
        \end{subfigure}
        \begin{subfigure}[h]{0.45\linewidth}
        \includegraphics[width=\linewidth]{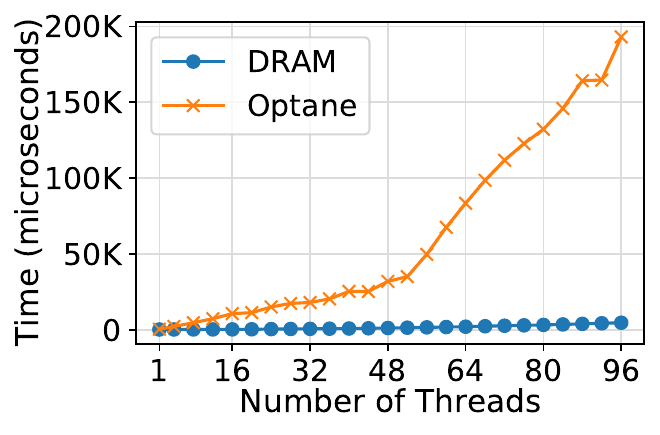}
        \caption{Page order 9}
        \label{fig:line_alloc_varth_or9}
        \end{subfigure}
    \caption{Multi-threaded scalability analysis of page allocation cost on DRAM and Optane for order-0 (4\,\text{KB}) and order-9 (2\,\text{MB}) pages. Note the difference in y-axis scale}
        \label{fig:line_alloc_varth}
    \end{figure}
We perform scalability analysis with order 0 (4\,\text{KB}) and order 9 (2\,\text{MB}) pages which are the hardware supported pages sizes on Intel systems as well as the most frequently allocated page orders. 
We allocate (150\,\text{GB}) of memory in total on a single DRAM or Optane node and bind the threads to CPU cores.

For the first 48 threads, page allocation is local to the socket. For threads beyond 48, page allocations is either local or remote to the socket depending on benchmark thread's affinity to the NUMA node.

We observe that average page allocation cost for Optane remains comparable to DRAM memory till 8 concurrent threads, but the cost starts increasing thereafter (Figure~\ref{fig:line_alloc_varth}).
Furthermore, page allocation cost increases drastically after 48 threads for Optane when the threads spill over to the second socket, however, such drastic increase after 48 threads is not observed for DRAM. 
Compared to single threaded allocation cost, the DRAM page allocation cost for 96 threads increases by 10$\times$ and 11.85$\times$ for order-0 and order-9, while for Optane,
the cost increases by 196$\times$ and 300$\times$. The performance gap between DRAM and Optane for 96 threads is 25$\times$ and 39.6$\times$ for order-0 and order-9 pages, respectively.
    
\subsubsection{Discussion}
The cost of allocating higher order pages on Optane,
    especially order-9 (2\,\text{MB}) pages that are frequently allocated to serve huge page allocation requests, is a concern. For terabyte scale memory technologies such as Optane, efficient allocation and management of 2\,\text{MB} huge pages is important in order to improve TLB efficiency.
    Therefore, it is essential to employ optimizations to scale order-9 allocation cost.

\subsection{Memory capacity scalability}
\label{char:memory-capacity-scalability}
\label{sec:mem_cap_scalability}
\begin{figure}
\centering
    \begin{subfigure}[h]{\linewidth}
    \centering
    \includegraphics[width=0.9\linewidth]{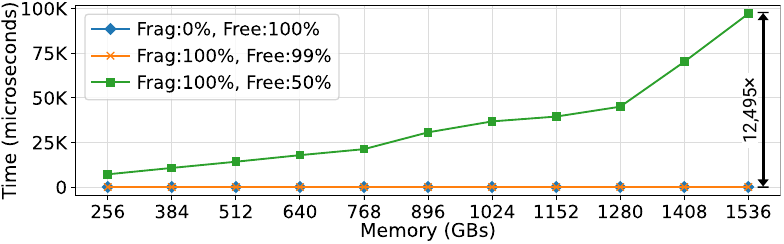}
    \caption{50\textsuperscript{th} percentile}
    \label{fig:line_frag_50}
    \end{subfigure}
    \begin{subfigure}[h]{\linewidth}
    \centering
    \includegraphics[width=0.9\linewidth]{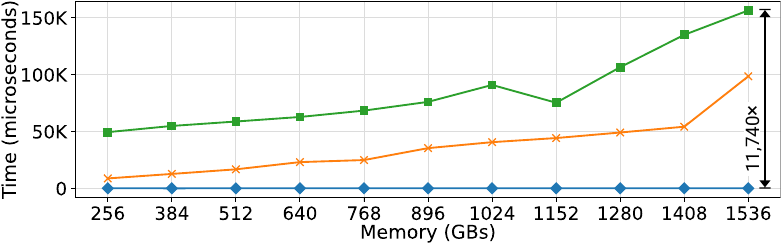}
    \caption{99\textsuperscript{th} percentile}
    \label{fig:line_frag_99}
    \end{subfigure}
    \caption{Memory capacity scalability analysis of single threaded order-9 page allocation on Optane with different fragmentation levels and free memory availability (x-axis represents available memory capacity after soft offlining the required number of memory blocks)}
    \label{fig:line_frag_all}
    \end{figure}

In this section we analyze the impact of memory capacity scalabality on page allocation for order-9 pages. We aim to understand the impact of fragmentation and free memory availability on the page allocation latency at terabyte scale.

Memory fragmentation is unavoidable in real-world environments, especially when allocation of multiple page orders are supported. In a fragmented system, memory needs to be defragmented in order to allocate large contiguous regions such as huge pages. 
To emulate a fragmented system, we completely fragment the Optane NUMA node to ensure no contiguous memory region of order-9 (2\,\text{MB}) is available to serve allocation requests in the fast path. 
On a completely fragmented Optane NUMA node, we ensure either 50\% or 99\% of the memory is free in every contiguous 2\,\text{MB} region by freeing the required number of 4\,\text{KB} pages as followed in prior works~\cite{ZhuSuperpageUsenix}.
For instance, in the scenario \texttt{Frag:100\%, Free:50\%}, not even a single contiguous 2\,MB region of memory is available and hence the system is 100\% fragmented. However, in each 2\,MB region, out of 512 available 4\,KB pages 256 4\,KB pages are free and hence 50\% is free. This implies that compaction should move 256 4\,KB pages to a different location to have a 2\,MB contiguous free region.

We perform the experiment for different memory capacity by suitably soft offlining the required number of memory blocks (using the \texttt{sysfs} interface). 
We allocate 64\,\text{GB} of order-9 pages with memory defragmentation or  compaction~\cite{memorycompaction} allowed from the page allocation context. To avoid interference caused by multi-threading and page zeroing, we use a single threaded benchmark and skip page zeroing. 

Figure~\ref{fig:line_frag_all} shows the impact of memory capacity on page allocation latency when Optane node is not fragmented (\texttt{Frag:0\%}) and fully fragmented (\texttt{Frag:100\%}). We observe that page allocation latency remains unaffected when the system is not fragmented and is thus neutral to the terabyte scale Optane memory capacity. With 100\% fragmentation and 99\% free memory, the median page allocation latency is unaffected, but the 99\textsuperscript{th} percentile allocation latency increases with the increase in memory capacity (7393$\times$ for 1536\,\text{GB}) compared to the unfragmented scenario.
However, with 100\% fragmentation and 50\% free memory, 
the page allocation cost increases significantly with increase in memory capacity.
For example, with 1536\,\text{GB} memory, the median and 99\textsuperscript{th} percentile allocation latencies are 12495$\times$ and 11740$\times$, respectively, compared to the unfragmented scenario. 

\subsubsection{Discussion}

Page allocation latency remains unaffected for larger memory capacity if the system is not fragmented. However, once the memory is fragmented, the median allocation latency increases sharply even when 50\% of the memory is free on Optane. The increase in allocation latency is due to the additional time required by the memory subsystem to scan, identify and compact memory pages. Hence, memory management subsystems should employ aggressive fragmentation avoidance techniques on systems that are equipped with terabyte scale memory tiers. 

\subsection{Huge page management} 
\label{char:transparent-huge-page}
\begin{table}
    \centering
    \footnotesize
    \begin{tabular}{ c c c c c }
        \toprule
        \textbf{Scan rate}  & \textbf{Unlimited} & \textbf{High} & \textbf{Moderate} & \textbf{Default} \\
        \midrule
        \textbf{Scan sleep (sec.)}  & 0 & 0.1 & 1 & 10 \\ 
        \hline
        \textbf{Promotion time}  & 26 mins. & 2.08 hrs. & 6.9 hrs. & 21.4 hrs. \\ \hline
    \end{tabular}
    \caption{Time taken to identify and promote 1.4\,\text{TB} of application's memory region to huge pages}
    \label{tab:khugepaged}
\end{table}

In this section, we aim to understand the performance of huge page managemet at terabyte scale. Deferred allocation avoids high huge page allocation latency during a page fault by initially allocating a 4\,\text{KB} page. The
\texttt{khugepaged} kernel daemon identifies and promotes a set of 512 contiguous 4\,\text{KB} pages to a huge page by
scanning the virtual address space of a process.

We allocate 1.4\,\text{TB} of memory from our single threaded userspace process using \texttt{mmap()} system call and then enable huge page promotions by invoking
\texttt{madvise(MADV\_HUGEPAGE)}. We perform the experiment on a non-fragmented memory and measure the promotion time for different scan rates. We set the number of pages to scan at each pass to 1 million. Sleep time between each pass controls the frequency and aggression of \texttt{khugepaged}. Lower the sleep time, higher is the promotion rate, but at the cost of higher CPU utilization. 

With unlimited scan rate (100\% utilization of one CPU core), \texttt{khugepaged} takes 26 minutes to scan and promote 1.4\,\text{TB} of memory, while the default settings on Linux
takes 21.4 hours (Table~\ref{tab:khugepaged}). High, moderate and default scan rates results in 5--48\% of CPU utilization in bursts at different intervals of time.

\subsubsection{Discussion}
Aggressive promotion of application's memory region to huge pages reduces TLB overheads which in turn improves throughput but at the cost of dedicating one CPU core for huge page management. It is important to note that huge page promotions
are not one time work. In fact scanning and promoting huge pages (with associated CPU overheads) are ongoing activities because applications can dynamically allocate and deallocate memory. In addition, huge pages can be split or demoted into multiple 4\,\text{KB} pages, for example, during compaction and NUMA rebalancing. They should be identified and promoted again when they are eligible for promotion.

\subsection{Performance tuning implications}   %/* BFS, Memcached */
\label{char:perform-tuning}
\begin{figure}
    \begin{subfigure}[h]{0.48\linewidth}
        \includegraphics[width=\linewidth]{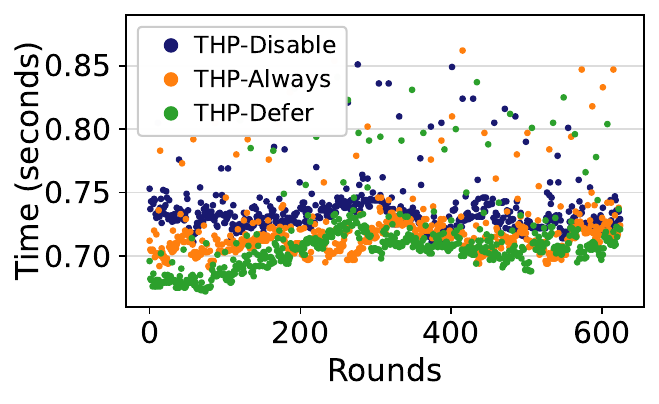}
        \caption{DRAM}
        \label{fig:scatter_BFS_DRAM}
    \end{subfigure}
    \begin{subfigure}[h]{0.48\linewidth}
        \includegraphics[width=\linewidth]{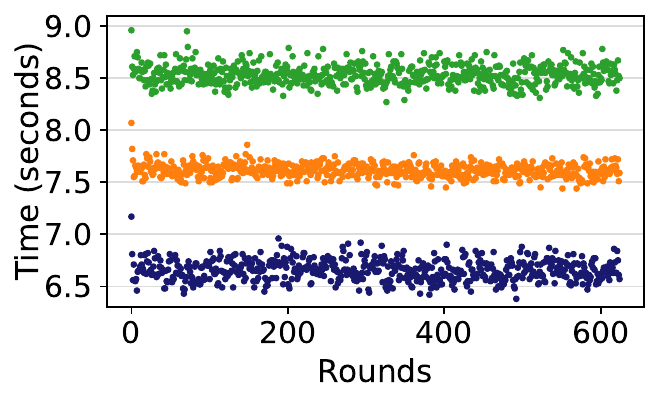}
        \caption{Optane}
        \label{fig:scatter_BFS_OPTANE}
    \end{subfigure}
    \caption{BFS graph traversal time (lower is better) with THP demonstrating how Optane inverts the prior DRAM-based optimizations and tuning. Y-axis is the traversal time for each round}
    \label{fig:scatter_BFS}
\end{figure}

\begin{figure}
\centering
        \includegraphics[scale=0.38]{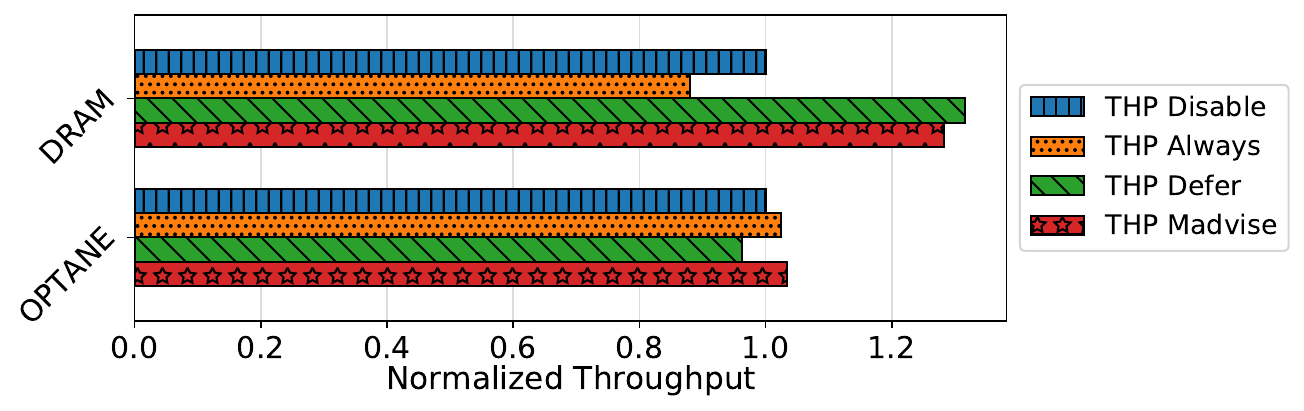}
        \caption{Performance comparison (normalized to THP Disable) of Memcached with different THP configurations}
        \label{fig:scatter_mcd}
\end{figure}

We demonstrate the impact of performance tuning on Optane memory with BFS graph traversal algorithm~\cite{bfs} and Memcached in-memory database~\cite{memcached}. In BFS we perform 625 rounds of traversal on an \texttt{rMat27} graph which allocates around 72\,\text{GB} of memory. For Memcached we use a 300\,\text{GB} in-memory database with 4\,\text{KB} key-value size and hotspot data set to 20\%. We use YCSB~\cite{ycsb} to generate the load.
We experiment with different THP settings by binding the memory of the applications to either DRAM or Optane memory.

As shown in Figure~\ref{fig:scatter_BFS}, on DRAM
BFS performs the best for \texttt{THP-Defer} (i.e., defer huge page allocation to \texttt{khugepaged} daemon) with some overlap with \texttt{THP-Always} (allocate huge page during page fault), while \texttt{THP-Disable} (allocation of huge pages disabled) performs the worst. However, with Optane memory, as expected, the time taken to complete a single round of graph traversal increases due to higher media access latency, but surprisingly \texttt{THP-Defer} that worked best on DRAM performs worst on Optane.
We observe similar behaviour with Memcached as well (Figure~\ref{fig:scatter_mcd}) where \texttt{THP-Defer} that worked best on DRAM performs worst on Optane. As Memcached supports application controlled huge page allocation (\texttt{THP-Madvise}) we include it in our comparison. 

We think that finding a 2\,\text{MB} contiguous region upon a fault in \texttt{THP-Always} is costly on DRAM as contiguous regions are not always available triggering compaction in the fault handler, but promoting a set of base pages to huge pages by a background daemon in \texttt{THP-Defer} is efficient.
However, on Optane, as contiguous regions are mostly available due to large memory capacity, compaction is not triggered in \texttt{THP-Always}. But due to large memory capacity, background daemon in \texttt{THP-Defer} requires more cycles to identify and promote pages to huge pages and hence is not efficient. However, we feel further deep dive into the problem is required to clearly identify the performance anomaly with huge pages on Optane.

\subsubsection{Discussion}

Data of large memory footprint applications is expected to reside in both DRAM and Optane in a tiered memory system. Hence, when a certain optimization or tuning is applied, a subset of application's data DRAM benefits out of the optimization while the other subset of data in Optane may not benefit or in the worst case suffers performance degradation (or vice versa). 
Therefore a system-wide optimization and tuning may not result in expected performance improvements. Thus memory management optimizations and tuning that can be targeted to specific memory tiers are required. In addition, applications require careful reevaluation of currently employed performance tuning for tiered memory systems.

\subsection{Impact of interference}
\label{char:factors-page-alloc}
In this section we analyze the impact of interference on page allocation cost by triggering concurrent page faults using a userspace benchmark process that allocates 3\,\text{GB} of memory per thread. We bind the benchmark process to Socket 0 and allocate memory from the local DRAM or Optane NUMA node. In parallel we start a multi-threaded interference process on Socket 1 where each thread continuously writes to a 256\,\text{MB} memory mapped region in a loop on DRAM or Optane nodes depending on where the benchmark is allocating memory from. 

\begin{figure}
\centering
        \includegraphics[width=\linewidth]{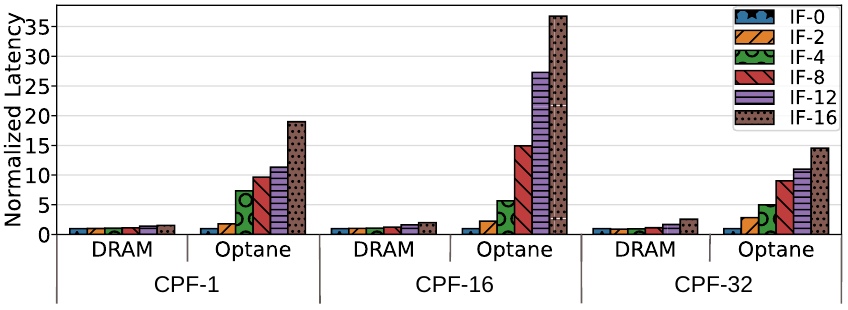}
        \caption{Impact of interference on page allocation cost. The latency values are normalized with respect to \texttt{IF-0}}
        \label{fig:bandwidth_interference}
\end{figure}

Figure~\ref{fig:bandwidth_interference} shows the impact of multi-threaded interference normalized to \texttt{IF-0} (\texttt{x} in \texttt{IF-x} represents the number of interference threads. \texttt{IF-0} represents no interference) on page allocations when concurrent page faults (CPF) are triggered (\texttt{y} in \texttt{CPF-y}, represents the number of concurrent faults). 
We observe that the page allocation cost remains within acceptable limits on DRAM while it increases significantly as we increase the number of interference threads on Optane.
For instance, on Optane with \texttt{CPF-1} (single threaded page fault), even with 2-threaded and 4-threaded interference process, the allocation cost increases by 1.79$\times$ and 7.34$\times$. It can be noted that in both the cases the system is almost idle with around 3\% and 5\% CPU utilization.

\subsubsection{Discussion}
Page allocation cost on Optane is sensitive to interference. Interference can be either from applications reading/writing to Optane memory or due to reads/writes triggered by page migration during NUMA rebalancing, memory compaction and hot/cold pages promotion/demotion. While it is difficult to avoid interference from applications, interference from memory management activities can be avoided by carefully coordinating/scheduling them.

\subsection{Virtualized environments}
 \label{char:virtual-env}
 \begin{figure}
 \centering
    \begin{subfigure}[h]{0.3\linewidth}
    \centering
       \includegraphics[width=\linewidth]{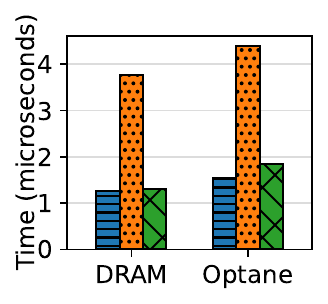}
        \caption{Page order 0}
        \label{fig:bar_alloc_vm_or0}
    \end{subfigure}
    \captionsetup[subfigure]{oneside,margin={-2.1cm,0cm}}
    \begin{subfigure}[h]{0.6\linewidth}
    \centering
        \includegraphics[width=\linewidth]{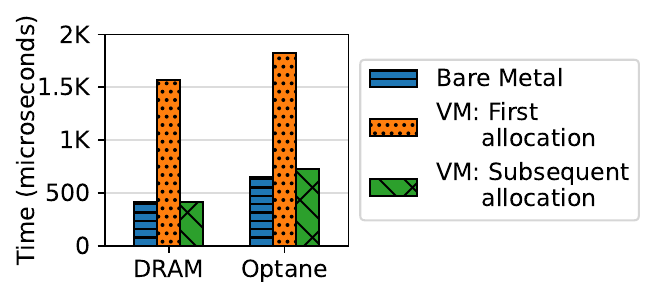}
        \caption{Page order 9}
        \label{fig:bar_alloc_vm_or9}
    \end{subfigure}
    \caption{Page allocation cost for the first and subsequent allocations in a virtualized environment on an idle system that highlights double zeroing issue}
    \label{fig:bar_alloc_vm}
\end{figure}

The page allocation overheads described in this section are also applicable to virtualized environment. However,
when a physical page is allocated to the VM for the first time, the page allocator in the host zeros the contents of the page which is immediately zeroed again by the page allocator in the guest OS. This results in page being zeroed twice~\cite{SilentShredder}. 
Figure~\ref{fig:bar_alloc_vm} shows the average page allocation cost on a \texttt{KVM/QEMU} VM with 32\,\text{GB} memory and one vCPU (to avoid multi-threaded interference). Due to double zeroing, the first page allocation cost inside the VM is more than 2$\times$ the cost of allocating a page on native system. But reallocating a page to a different process within the guest do not incur double zeroing as the host is not involved.

\subsubsection{Discussion}

Double zeroing in VMs is a known issue~\cite{SilentShredder}, however, we note that the double zeroing problem is exacerbated when page is allocated on Optane, especially when other factors as discussed above such as concurrent page faults, allocation locality, fragmentation are also in play. 

Furthermore, on a memory over-committed host, which is a norm in enterprise cloud environments, balloon drivers can frequently reassign memory across multiple guests. Allocating a page from reassigned memory also incurs double zeroing overheads.
Hence, new set of interfaces between host and guest to eliminate double zeroing where the host communicates to the guest that it has already zeroed the page so that the guest can skip page zeroing is worth exploring. 

\section{Summary and Discussion}
\label{sec:discussions}
\subsection{Zeroing overheads}

\begin{table}
    \centering
    % \footnotesize
    \begin{tabular}{lrr}
        \toprule
        \textbf{Workload} & \textbf{DRAM} & \textbf{Optane} \\
        \midrule
        BFS  & 3.14\% & 16.29\%   \\ \hline
        Firecracker VM (multi instance boot) & 5.15\% & 20.44\% \\ \hline
        Node.js (multi instance run) & 1.40\% & 19.38\% \\ \hline
		Memcached (DB init) & 36.43\% & 52.31\% \\ \hline
		%Kernel Build & 1.97\% & 4.49\% \\ \hline
    \end{tabular}
    \caption{Percentage of total application execution time spent in zeroing during page allocation on DRAM and Optane}
    \label{tab:zero_percent}
\end{table}

Memory allocation and deallocation costs significantly contribute to ``datacenter tax'' 
(cycles spent in low level software stack including the OS kernel)~\cite{sriraman2020accelerometer, 10.1145/2749469.2750392}.
As observed in \S\ref{char:page_alloc_overheads}, page zeroing overhead with Optane can be a significant part of the datacenter tax.
Our analysis of real-world benchmarks (Table~\ref{tab:zero_percent}) shows that
around 16.29--52.31\% of total execution time is spent on zeroing for Optane, while it is 1.4--36.43\% for DRAM.

Prior works have looked into optimizing zeroing cost on DRAM-based memory~\cite{hawkeye, intel_arc_opt_reference}, however with fundamentally different hardware properties, such optimizations can be ineffective on Optane memory.
The Enhanced REP MOVSB and STOSB (ERMS)~\cite{intel_arc_opt_reference} operation supported on Intel Architecture and exploited in the Linux kernel still results in high zeroing overhead on Optane.
Optimizations such as asynchronous or background zeroing as well as multi-threaded zeroing, effective for DRAM, can saturate the limited memory bandwidth on Optane and impact the application performance. 

We believe new techniques, designed to eliminate page zeroing from the OS page allocator, can help in leveraging Optane's capacity advantages. 
Techniques such as (i) hardware-software co-design, for example, in-memory processing
that offloads zeroing to NVDIMMs, (ii) applications or compilers providing crucial hints 
to the OS to skip zeroing, (iii) zeroing elimination through obfuscation where the data is rendered difficult or impossible to interpret by changing few data bits~\cite{SilentShredder} and (iv) maintaining process affinity of freed pages to enable reallocating the page to the same process without zeroing are worth exploring in this context. 

However, for allocation instances where zeroing cannot be eliminated, the OS memory management subsystem should consider optimizing Optane induced performance bottlenecks such as high
cross socket zeroing cost (\S\ref{char:allocation-locality}) by employing techniques that
offload zeroing to the remote socket.

\subsection{Coordinated memory management}

The DRAM based core kernel memory management functionalities employ strategies and heuristics that are restricted to core memory attributes such as free memory thresholds, memory fragmentation index, but do not consider other system attributes such as memory bandwidth utilization, page fault rate. 
For example, heuristics employed in AutoNUMA do not consider the page allocation rate when triggering page migrations. However, page migrations can cause noticeable interference on Optane NUMA nodes because the bandwidth consumed during memory move operations can result in drastic increase in allocation latency.
Unlike DRAM, memory management functionalities such as page allocation on Optane are sensitive to other system activities and thus requires carefully coordinated techniques across multiple memory management subsystems.

\subsection{Optimizing huge page allocations}
For terabyte scale memory, allocating and managing 2\,\text{MB} huge pages are important.
We believe that it is essential to employ optimizations that improve huge page performance, for example, by extending the per-cpu list of free pages to order-9 pages as well.
We also observe that on-fault allocation of order-9 pages on Optane incurs a high allocation latency, especially when large capacity memory is fragmented.
In such scenarios, on-fault allocations are not preferable. Linux allocates huge pages either on-fault or defer the allocation depending on the system-wide static THP setting. 
For Optane, on-fault or deferred allocation can be dynamic 
based on the memory fragmentation levels.

A huge page can be split into multiple 4\,\text{KB} pages, for example, during compaction and NUMA rebalancing. It is important to aggressively identify and promote them again to huge pages as soon as they are eligible for promotion. 
Existing single threaded implementation of \texttt{khugepaged} daemon requires several minutes to scan, identify and promote eligible huge page regions for systems with terabyte memory capacity. Implementing multi-threaded \texttt{khugepaged} daemon may not be viable as it can saturate memory bandwidth due to page migrations during parallel huge page promotions.
Hence, exploring novel huge page memory management techniques for terabyte scale Optane memory is essential.

\subsection{Fragmentation avoidance}
Fragmented Optane memory increases median allocation latency by 12,495$\times$ compared to unfragmented scenario.
Even though the Linux kernel employs several heuristics in page allocation, page reclamation and other functionalities to avoid fragmentation, in some cases it prefers local allocation over fragmentation avoidance, i.e., prefers allocating a page on local NUMA node instead of a remote NUMA node even when the allocation can potentially fragment the memory. 
This is a fair trade-off in the DRAM memory context because local memory accesses
improve application performance and fragmentation has limited impact on the allocation efficiency.
Further, the defragmentation cost, a function of memory capacity, is not costly for DRAM. However, on Optane both local allocation and fragmentation avoidance are important because remote accesses are costly and fragmentation exorbitantly increases median allocation latency.
Moreover, defragmenting large capacity Optane consumes significant amount of CPU cycles and can stress the limited memory bandwidth available on Optane nodes.

Hence, memory management techniques prioritizing both local allocation and fragmentation avoidance should be investigated for Optane memory systems.

\subsection{Differential memory management}
Due to fundamental differences in media properties and memory capacity of DRAM and Optane, the optimizations and configurations suitable for DRAM may not work for Optane memory (\S\ref{char:perform-tuning}).
However, on a tiered memory system the memory footprint of large applications can span both DRAM and Optane. But the existing OSes support a system-wide tuning and optimization options which cannot be applied specific to DRAM and Optane memory. 
As a result, these optimizations and tuning work best only for datasets in either DRAM or Optane memory.
In addition, Optane memory requires heuristics and strategies that are different from
DRAM (e.g., fragmentation avoidance, coordinated management as discussed above). A global system-wide memory management scheme may not perform the best on both DRAM and Optane memory tiers.
 
Therefore, OSes should provide differential memory management for different memory tiers.
However, incorporating differential management into complex memory management subsystems on production-grade OSes is difficult and may require significant design changes or a rethink of existing memory management principles.

\section{Related Work}
\subsection{Performance Analysis} Yang et.al.~\cite{fastpaper} evaluates Optane performance with different data access patterns and provides best practices for application developers on using system with Optane persistent memory.
Performance characterization of different applications such as graph analytics, Java workloads, and high-performance computing have also been evaluated on Intel Optane systems~\cite{optane_graph,10.1145/3451342,hpc_optane_measurement}. 

0sim~\cite{0simMark} analyzes memory management overheads on terabyte-scale NVMMs using a novel software simulator (real hardware was not available at that time). However, as observed in~\cite{fastpaper} and also in our work, performance numbers captured on NVMM simulators/emulators are widely inaccurate compared to the real hardware.

Xiang et.al.~\cite{Xiang_onDIMM_Eurosys22} analyze the performance of on-DIMM buffering in Optane and highlights the performance impact using microbenchmarks.

To the best of our knowledge, this is the first work that analyzes and studies the effectiveness of physical memory management for Optane in detail. 
\subsection{System software optimizations for NVMM} Prior works have shown that efficient placement and migration of pages across memory tiers have an important role in overall performance~\cite{nimble,mitosis,kumar2021radiant,HeMem21,TMO_asplos22,pond_asplos23,TPP_asplos23}.
Nimble page management~\cite{nimble} identified various bottlenecks associated with page management and migration on a tiered memory system.
Radiant~\cite{kumar2021radiant} proposes efficient page table placement policies in hybrid memory systems.
FairHym~\cite{FairHymNVMSA} works on improving inter-process fairness on hybrid memory system to prevent performance degradation while making use of both media.

While most of the existing research propose interesting optimizations to different memory subsystems, this work focuses on providing useful insights related to physical memory management aspects of the OS.    

\subsection{Software abstractions for efficient use of NVMM} Several system software abstractions (new and augmented) for efficient access of NVMM are proposed ~\cite{Bittmantwizzler, splitfs,bpfs2009,dulloorNVMsystemsoftware,socc20remotepmm,remotePM2}.
Twizzler~\cite{Bittmantwizzler} proposes an OS with persistent object support.  
Specialized file systems such as SplitFS~\cite{splitfs} provide efficient POSIX file semantics using NVMM.
Researchers have explored redesigning modern data centers to provide networked access to NVMM from remote servers~\cite{socc20remotepmm,remotePM2}.
For all the solutions mentioned above, efficient management of physical memory on NVMM plays a crucial role.

\section{Conclusions}
The DRAM-NVMM hybrid memory system promises a cost effective, scalable solution to the ever expanding memory requirements of data centers.
From our comprehensive empirical analysis of physical memory management in hybrid memory systems, we observed that several existing mechanisms and optimizations for DRAM fall short for NVMM, primarily due to the difference in hardware characteristics. 
We highlighted the problems and potential solution directions 
to motivate the need to develop novel memory
management techniques for terabyte scale NVMMs.

{\footnotesize
\bibliographystyle{plain}
\bibliography{references}}

\end{document}